%Paper: hep-th/9302002
%From: eabdalla%uspif.hepnet@Lbl.Gov
%Date: Mon, 1 Feb 93 09:08:57 PST

\magnification 1200
%\input macrosbk.tex
%\magnification = 1200
%

%
\font\eightrm=cmr8
\font\eighti=cmmi8
\font\eightsy=cmsy8
\font\eightbf=cmbx8
\font\eighttt=cmtt8
\font\eightit=cmti8
\font\eightsl=cmsl8
\font\sixrm=cmr6
\font\sixi=cmmi6
\font\sixsy=cmsy6
\font\sixbf=cmbx6
\catcode`@11
\newskip\ttglue
\font\grrm=cmbx10 scaled 1200

\def\eightpoint{\def\rm{\fam0\eightrm}
\textfont0=\eightrm \scriptfont0=\sixrm \scriptscriptfont0=\fiverm
\textfont1=\eighti \scriptfont1=\sixi \scriptscriptfont1=\fivei
\textfont2=\eightsy \scriptfont2=\sixsy \scriptscriptfont2=\fivesy
\textfont3=\tenex \scriptfont3=\tenex \scriptscriptfont3=\tenex
\textfont\itfam=\eightit \def\it{\fam\itfam\eightit}
\textfont\slfam=\eightsl \def\sl{\fam\slfam\eightsl}
\textfont\ttfam=\eighttt \def\tt{\fam\ttfam\eighttt}
\textfont\bffam=\eightbf
\scriptfont\bffam=\sixbf
\scriptscriptfont\bffam=\fivebf \def\bf{\fam\bffam\eightbf}
\tt \ttglue=.5em plus.25em minus.15em
\normalbaselineskip=6pt
\setbox\strutbox=\hbox{\vrule height7pt width0pt depth2pt}
\let\sc=\sixrm \let\big=\eightbig \normalbaselines\rm}
\newinsert\footins
\def\newfoot#1{\let\@sf\empty
  \ifhmode\edef\@sf{\spacefactor\the\spacefactor}\fi
  #1\@sf\vfootnote{#1}}
\def\vfootnote#1{\insert\footins\bgroup\eightpoint
  \interlinepenalty\interfootnotelinepenalty
  \splittopskip\ht\strutbox % top baseline for broken footnotes
  \splitmaxdepth\dp\strutbox \floatingpenalty\@MM
  \leftskip\z@skip \rightskip\z@skip
  \textindent{#1}\footstrut\futurelet\next\fo@t}
\def\fo@t{\ifcat\bgroup\noexpand\next \let\next\f@@t
  \else\let\next\f@t\fi \next}
\def\f@@t{\bgroup\aftergroup\@foot\let\next}
\def\f@t#1{#1\@foot}
\def\@foot{\strut\egroup}
\def\footstrut{\vbox to\splittopskip{}}
\skip\footins=\bigskipamount % space added when footnote is present
\count\footins=1000 % footnote magnification factor (1 to 1)
\dimen\footins=8in % maximum footnotes per page

\def\ref#1{$^{#1}$}
\def\flex{\raise 6pt\hbox{$\leftrightarrow $}\! \! \! \! \! \! }
\def\oversome#1{ \raise 8pt\hbox{$\scriptscriptstyle #1$}\! \! \! \! \! \! }

\newbox\bigstrutbox
\setbox\bigstrutbox=\hbox{\vrule height10pt depth5pt width0pt}
\def\bigstrut{\relax\ifmmode\copy\bigstrutbox\else\unhcopy\bigstrutbox\fi}
\def\refer[#1/#2]{ \item{#1} {{#2}} }
\def\rev<#1/#2/#3/#4>{{\it #1\/} {\bf#2}, {#3}({#4})}
\def\boxit#1{\vbox{\hrule\hbox{\vrule\kern3pt
\vbox{\kern3pt#1\kern3pt}\kern3pt\vrule}\hrule}}

\def\2figure#1#2#3#4{\vbox{ \hrule width#1truecm \hbox{\vrule height#2truecm
\hskip #1truecm
\vrule height#2truecm }\hrule width#1truecm \hbox{\vrule\vbox{\hsize #1truecm
\baselineskip=10pt
\noindent\strut#3}\vrule}\hrule width#1truecm
\hbox{\vrule\vbox{\hsize #1truecm
\baselineskip=10pt
\noindent\strut#4}\vrule}\hrule width#1truecm  }}
\def\3figure#1#2#3#4#5{\vbox{ \hrule width#1truecm \hbox{\vrule height#2truecm
\hskip #1truecm
\vrule height#2truecm }\hrule width#1truecm \hbox{\vrule\vbox{\hsize #1truecm
\baselineskip=10pt
\noindent\strut#3}\vrule}\hrule width#1truecm
 \hbox{\vrule\vbox{\hsize #1truecm
\baselineskip=10pt
\noindent\strut#4}\vrule}
\hrule width#1truecm \hbox{\vrule\vbox{\hsize #1truecm
\baselineskip=10pt
\noindent\strut#5}\vrule}\hrule width#1truecm  }}

\def\sqr#1#2{{\vcenter{\hrule height.#2pt
   \hbox{\vrule width.#2pt height#1pt \kern#1pt
    \vrule width.#2pt}
    \hrule height.#2pt}}}

% Here are my additional definitions:

\def\smin{\,\raise 0.06em \hbox{${\scriptstyle \in}$}\,}
\def\smsubset{\,\raise 0.06em \hbox{${\scriptstyle \subset}$}\,}

\def\Natural{\hbox{\hskip 1.5pt\hbox to 0pt{\hskip -2pt I\hss}N}}

\def\Rational{\hbox{\hbox to 0pt{\hskip 2.7pt \vrule height 6.5pt
                                  depth -0.2pt width 0.8pt \hss}Q}}
\def\Real{\hbox{\hskip 1.5pt\hbox to 0pt{\hskip -2pt I\hss}R}}
\def\Complex{\hbox{\hbox to 0pt{\hskip 2.7pt \vrule height 6.5pt
                                  depth -0.2pt width 0.8pt \hss}C}}


\vfill\eject
\nopagenumbers
\centerline {\grrm Off critical current algebras}
\vskip 1.5cm
\centerline {\bf E. Abdalla\ref{1}, M.C.B. Abdalla\ref 2, G.
Sotkov\ref{2}\newfoot{$^*$}{On leave of absence from Bulgarian Academy of
Science, Sofia}, M. Stanishkov\ref 1.}
\vskip .5cm
\centerline { Instituto de F\'\i sica da Universidade de S\~ao Paulo\ref 1}
\centerline { C.P. 20516, S\~ao Paulo, Brazil.}
\centerline { Instituto de F\'\i sica da Universidade Estadual Paulista, Julio
de Mesquita\ref 2}
\centerline{ Rua Pamplona, 145, CEP 01405, S\~ao Paulo, Brazil}
\vskip 2cm
\centerline {\bf Abstract}
\vskip.4cm
We discuss the infinite dimensional algebras appearing in integrable
perturbations of conformally invariant theories, with special emphasis in the
structure of the consequent non-abelian infinite dimensional algebra
generalizing  $W_\infty$ to the case of a non abelian group. We prove that the
pure left-symmetry  as well as the pure right-sector of the thus obtained
algebra coincides with the conformally invariant case. The mixed sector is more
involved, although the general structure seems to be near to be unraveled. We
also find some subalgebras that correspond to Kac-Moody algebras. The
constraints imposed by the algebras are very strong, and in the case of the
massive deformation of a non-abelian fermionic model, the symmetry alone is
enough to fix the 2- and 3-point functions of the theory.
\vskip.5cm

\hfill Universidade de S\~ao Paulo\quad

\hfill IFUSP-preprint-1027\phantom{Paulo}\quad

\hfill January 1993\phantom{Paulo}\quad

\vfill
%{\eightrm This work has been supported by a FAPESP award.}
\eject

\countdef\pageno=0 \pageno=1
\newtoks\footline \footline={\hss\tenrm\folio\hss}
\def\folio{\ifnum\pageno<0 \romannumeral-\pageno \else\number\pageno \fi}
\def\advancepageno{\ifnum\pageno<0 \global\advance\pageno by -1
\else\global\advance\pageno by 1 \fi}

\centerline {{\bf 1.} Introduction}
\vskip .5cm

The appearance of the Virasoro\ref{1} and the $W_\infty(V)$ algebras\ref{2} as
off-critical symmetries of certain class of integrable models (IM) addresses
the
question whether all the known infinite algebras (super-Virasoro, $\widehat
G_n$-Kac-Moody, $W_N$, etc $\cdots$) reappear again as
symmetries of the {\it nonconformal} integrable models. From the way Virasoro
algebra
arises in the off-critical $XY$, Ising and Potts models one could expect to
find
the $\widehat G_n$-Kac-Moody algebra (and the larger current algebra $\widehat
G_n\subset\!\!\!\!\!\!\times {\rm Vir}$) studying the algebra of the conserved
charges of the integrable perturbations\ref{3} of the conformal WZW
models\ref{4,10}.

The simplest example is the $k=2$, $SU(2)$-WZW model perturbed by the field
$\psi_{\Delta,j}^k(z) \overline \psi_{\Delta,j}^k(\overline z)$ of spin $j=1$
and
$\Delta =1/2$. This model is equivalent to the theory of massive Majorana
fermion in the $O(3)$-vector representation. A more general fact is that all
$k=1$, $O(n)$-WZW models represent free fermions\ref{5}, and their massive
perturbation is described by the action
$$
S=\int {1\over 2}\left( i\overline \psi^i \not \! \partial \psi^i + m\overline
\psi ^i\psi^i\right) d^2z\quad .\eqno(1.1)
$$

Our problem is to find all the symmetries of this simple fermionic system
and to see {\it whether the $\widehat O(n)$-Kac-Moody algebra (or
$\widehat O(n)\subset\!\!\!\!\!\!\times  {\rm Vir}$)-
algebra take place as symmetries of (1.1)}. The basic ingredients in the
construction of all the conserved charges for (1.1) are the following two
infinite sets of conserved tensors:
$$
\eqalign{T^{ji}_{2s}=T^{ij}_{2s}&=\psi^{(i}\partial ^{2s-1}\psi^{j)}\quad
,\quad \overline \partial T^{ij}_{2s}=\partial \theta ^{ij}_{2s-2}\cr
J^{ji}_{2s-1}=-J^{ij}_{2s-1} &= {1\over 2}\psi^{[i}\partial ^{2s-2}\psi^{j]}
\quad , \quad
\overline \partial J^{ij}_{2s-1}=\partial \widetilde \theta ^{ij}_{2s-3}\quad ;
s=1,2,\cdots\cr}\eqno(1.2)
$$

The conserved charges we are looking for are specific combinations of the
``higher momenta" in $z$ and $\overline z$ of these tensors. The first
surprising fact is that the charges of the currents $J^{ij}_{2s-1}\, , \,
\overline J^{ij}_{2s-1}$, namely
$$
\eqalign{Q^{ij}_s&= \int J^{ij}_{2s-1}dz - \int \widetilde
\theta^{ij}_{2s-3}d\overline z\cr
\overline Q^{ij}_s&= \int \overline J^{ij}_{2s-1}d\overline z - \int \widetilde
\theta^{ij}_{2s-3}dz\quad \quad s=1,2,\cdots\cr}
$$
close the ({\it nonchiral!}) $\widehat O(n)$-Kac-Moody algebra
$$
\left[ Q_{s_1}^{ij},Q_{s_2}^{kl}\right] =\delta_{ik}Q^{jl}_{s_1+s_2}+
\delta_{jl}Q^{ik}_{s_1+s_2} - \delta_{il}Q^{jk}_{s_1+s_2} -
\delta_{jk}Q^{il}_{s_1+s_2} + {p\over
2}s_1(\delta_{ik}\delta_{jl}-\delta_{il}\delta_{jk})\delta_{s_1+s_2}
\eqno(1.3)
$$
$(-\infty \le s_i\le \infty)$. The {\it central charge is $p=1$, i.e. the same
as for free massless fermions}. Constructing specific ``first momenta" of
$\delta_{ij}T^{ij}_{2s}$ (see eq.(3.13)) one can complete (1.3) to the larger
current algebra ${\rm Vir}\,\subset\!\!\!\!\!\!\times  \widehat O(n)$. The
appearance of this algebra as symmetries of (1.1) is a strong indication that
we can use their degenerate representations to solve exactly the model. As we
demonstrate in section 3 the simplest Ward identity of ${\rm Vir}\subset\!\!\!
\!\!\!\times  \widehat O(n)$ written for the fermionic two point function leads
to the $K_1$-Bessel equation.

It is important to mention that the $\widehat O(n)\,\subset\!\!\!\!\!\!\times
{\rm Vir}$-symmetry of the action (1.1) {\it have nothing to do} with the {\it
local gauge transformations } and {\it reparametrizations} in $2-D$
{\it ``space-time" $(z\overline z)$}. This is evident from the explicit higher
derivatives realization (3.9) and (3.13) of the action of the generators
$Q_s^{ij}$ and $V_k$ on $\psi^k(z,\overline z)$. The same formulae (3.9),
(3.13) written however in the $p$-space ($p\overline p=m^2)$ make transparent
the following {\it remarkable fact}: the algebra ${\rm Vir}
\subset\!\!\!\!\!\!\times  \widehat O(n)$ does represent specific local gauge
transformations and
reparametrizations in the $p$-space formulation of the model (1.1). The
parameters of these transformations $\omega^{ij}(p)$ and $\epsilon (p)$ have to
satisfy the conditions:
$$
\omega^{ij}(p)=\omega^{ij}(-p)\quad ,\quad \epsilon (p)=-\epsilon (-p)\quad .
$$
One should look for a better understanding of this fact in the {\it specific
phase space geometry of the model (1.1)}.

The current algebra ${\rm Vir}\subset\!\!\!\!\!\!\times  \widehat O(n)$ does
not
exhaust all the symmetries of the model (1.1). Together with the ``first
momenta" of $T^{ij}_{2s}$ and $J^{ij}_{2s-1}$ one can consider all the
conserved ``higher momenta" $\oversome{(-\!\!-)}\!L^{\,\, \, ij(2s)}_{-k}$ and
$\oversome{(-\!\!-)}\!Q^{\, \, ij(2s-1)}_{-p}\, (0\le k\le 2s-1\, ,\, 0\le p\le
2s-2)$ given by eqs.(3.17) and (3.18). The problem of deriving their algebra
requires some preliminary investigation. Having in mind the essential role of
the conformal $W_\infty$ algebra\ref{6} in the construction of the full
off-critical symmetry algebra  for the Ising model\ref{2} we have
first to find an appropriate
{\it conformal current algebra analog} of $W_\infty$. Our starting point is the
conformal OPE algebra of {\it all the descendents} $T^{ij}_{2s}$ of the
stress-tensor $T$
and $J^{ij}_{2s-1}$, of the $O(n)$-current $J^{ij}$. Following the standard
conformal technique we have constructed  in section 2
a {\it new class of $W_\infty$-current algebras} $W_\infty(\widehat G_n)$ for
$\widehat
G_n=\widehat O(n)$. We have calculated the structure constants $g_r^{s_1s_2}$
(2.7)
of $W_\infty(\widehat G_n)$  using the basis (2.3) of the quasiprimary
descendents.
This makes difficult the comparison with the corresponding structure constants
$c_r^{s_1s_2}$ of usual $W_\infty$\ref{6}, which are calculated in a specific
nonquasiprimary basis. There are however many indications that our universal
structure constants (2.7) has to coincide with these ones of $W_{1+\infty}$
[9],
taken in our quasiprimary basis.

The {\it off-critical analog} of the conformal $W_\infty(\widehat G_n)$ is
defined
as the algebra $\widetilde W_\infty (\widehat G_n)$ of all the
symmetries of (1.1). As
one can see from our discussion in section 4 it has quite complicated
structure: two subalgebras $\widetilde W_\infty ^{L(R)}(\widehat G_n)$ of
$W_\infty(\widehat G_n)$, which do not commute, two more
(incomplete $n\ge -1$) ${\rm Vir}\subset \!\!\!\!\!\!\times
\widehat O(n)$-current algebras, one $\widehat O(n)$-Kac-Moody algebra etc.
Similarly
to the simplest ${\rm Vir}\subset \!\!\!\!\!\!\times
\widehat O(n)$-algebra (1.3),  the generators displayed explicitly in (3.14)
have a specific higher
derivatives realization and {\it do not represent} local gauge transformations
and reparametrizations. For example, the generators of one of the incomplete
${\rm Vir}\subset\!\!\!\!\!\!\times
\widehat O(n)$-algebra are of the form:
$$
({\cal Q}^{ij}_n)_{kl}=-{1\over 2}(\delta^{ik}\delta^{jl} -
\delta^{il}\delta^{jk})[\overline z\overline \partial -z\partial -1]_n\partial
^n\quad ,\quad n\ge 0\quad ,\eqno(1.4)
$$
where $[A]_n=A(A-1)\cdots(A-n+1)$. The realization (1.4) is completely
different from the one of the {\it standard} $\widehat O(n)$-Kac-Moody
subalgebra
of the conformal $W_\infty(\widehat G_n) $:
$$
(Q^{ij}_n)_{kl}=-{1\over 2}(\delta^{ik}\delta^{jl} -
\delta^{il}\delta^{jk})z^n\quad ,\quad -\infty\le n \le \infty \quad
.\eqno(1.5)
$$

However the symmetries generated by (1.4) are not a specific feature of the
massive action (1.1) only. For the massless case where the full algebra of
symmetries is the conformal $W_\infty (\widehat G_n)$ one can find a subalgebra
spanned by
$$
(Q^{ij}_n)_{kl}=-{1\over 2}(\delta^{ik}\delta^{jl} -
\delta^{il}\delta^{jk})(z\partial +1)_n \partial ^n\eqno(1.6)
$$
which indeed close $\widehat O(n)$-Kac-Moody algebra (1.3).

All these similarities between the conformal $W_\infty (\widehat G_n)$ and the
off-critical $\widetilde W_\infty(\widehat G_n)$ reflect on the form of the
corresponding Ward identities as well. The conclusion is that in order to
construct the representations of $\widetilde W_\infty (\widehat G_n)$
and to solve its
Ward identities it is better to do it first for the conformal $W_\infty (
\widehat
G_n)$. In fact, the major part of the properties of the off-critical model are
hidden in the $W_\infty(\widehat G_n)$-symmetries of the conformal model we
have
started with.

\vskip 1.5cm
\penalty-400

\centerline{{\bf 2.} Conformal $W_\infty (\widehat G_n)$ algebras.}
\nobreak
\vskip .5cm
\nobreak
We are going to study the off-critical properties of certain integrable
perturbations of the  WZW models. The crucial role of
the conformal $W_\infty$ algebra in the construction of the corresponding
off-critical algebra $W_\infty (V)$ for the Ising and Potts models\ref{2}
addresses the question about the {\it relevant conformal current algebra
analog of $W_\infty$}. In the same way  $W_\infty$ algebra\ref{6,9} is
realized as the algebra of all the descendents
  $T_{2s}(z)$ $(s=1,2,\dots)$ of the
stress-tensor $T$ one expects the $W_\infty$-current algebra we are looking
for to be
spanned by the descendents $T_{2s}(z)$ and $J_{2s-1}^i(z)$, (or
$J_{2s-1}^{ij}={1\over 2}\varepsilon^{ijk}J_{2s-1}^k(z)$) of $T(z)$ and of the
$SU(2)$-currents $J^i(z)$. In order to find the $W_\infty (A_1)$ algebra we
have first to write the conformal OPE's of $J_{2s-1}^i$ and $T_{2s}$. We have
to be sure, however,  that $J^i_{2s-1}$, $T_{2s}$ form a basis in the space of
all the conserved tensors, i.e., the product of each two of them contains  the
currents from the defined basis only. The problematic OPE is the one  of
 two currents: $J^i_{2s-1}(z)J^k_{2p-1}(w)$. Its anti-symmetric
part can be exhausted by terms of the type $\varepsilon^{ijk}J^k_{2l-1}(z)$.
The
only remaining possibility for the symmetric part would be
$\delta^{ij}T_{2m}(z)$. This is however not generally the case, and one can see
this fact realizing $T_{2s}$ and $J^i_{2s-1}$ in terms of free fermions
$\psi^i$, $(i=1,2,3)$ as
$$\eqalign{J^i_{2s-1}&=\varepsilon^{ijk}\left( \partial^{2s-2}\psi^j\right)\psi
^k\cr
T_{2s}&=\psi^i\partial^{2s-1}\psi^i\quad . \cr} s=1,2,\cdots \eqno(2.1)$$
Using the standard OPE's for the fermions $\psi^i$ one can easily verify that
$\delta^{ij}T_{2s}$ does not exhaust the symmetric part of the aforementioned
OPE's. Indeed, there are new terms of the form
$$T^{ij}_{2s}=T^{ji}_{2s}= \psi^i\partial^{2s-1}\psi^j +\psi^j\partial^{2s-1}
\psi^i\quad , \quad T_{2s}={1\over 2} \delta^{ij}T_{2s}^{ij}\quad .$$
The conclusion is that the algebra we have to consider is the one of
$T^{ij}_{2s}$ and $J^{ij}_{2s-1}$. Applying the usual conformal
technique\ref{7,4}, one can easily derive the OPE's of the {\it
quasiprimary currents} $T^{ij}_{2s}$ and $J^{ij}_{2s-1}$. At this point we are
going to follow closely the method described in detail in section (3.5) of
ref.[8]. As a result of a simple and standard computation we obtain the
following OPE's algebra:
$$\displaylines{
T^{ij}_{2s_1}(z_1)T^{kl}_{2s_2}(z_2) =\sum_{s=1}\sum_{r=0}\Biggl\{
D_{2s}^{2s_1,2s_2}C_{2s}^{2s_1,2s_2}(r) z_{12}^{2(s-s_1-s_2)+r}\times\hfill\cr
\hfill\times\partial _2^r\Bigl(
\delta^{ik}T^{jl}_{2s}(2) + \delta^{il}T^{jk}_{2s}(2) +
\delta^{jk}T^{il}_{2s}(2) + \delta^{jl}T^{ik}_{2s}(2)\Bigr) \cr
+
D_{2s-1}^{2s_1,2s_2}C_{2s-1}^{2s_1,2s_2}(r) z_{12}^{2(s-s_1-s_2)+r-1}
\times\hfill\cr
\hfill\times\partial _2^r\Bigl(
\delta^{ik}J^{jl}_{2s-1}(2) + \delta^{il}J^{jk}_{2s-1}(2) +
\delta^{jk}J^{il}_{2s-1}(2) + \delta^{jl}J^{ik}_{2s-1}(2)\Bigr)\Biggr\} \cr
\hfill + N_{2s_1}^{(2)}z_{12}^{-4s_1}\delta_{s_1s_2}(\delta^{ik}\delta^{jl} +
\delta^{il}\delta^{jk})\hfill\cr
\noalign{\vskip 7pt}
T^{ij}_{2s_1}(z_1)J^{kl}_{2s_2-1}(z_2)=\sum_{s=1}\sum_{r=0}\hfill\cr
\hfill\Biggl\{
D_{2s}^{2s_1,2s_2-1}C_{2s}^{2s_1,2s_2-1}(r) z_{12}^{2(s-s_1-s_2)+r+1}
\partial^r\Bigl(\delta^{ik}T^{jl}_{2s}(2) - \delta^{il}T^{jk}_{2s}(2) +
\delta^{jk}T^{il}_{2s}(2) - \delta^{jl}T^{ik}_{2s}(2)\Bigr) \cr
+ D_{2s-1}^{2s_1,2s_2-1}C_{2s-1}^{2s_1,2s_2-1}(r)
z_{12}^{2(s-s_1-s_2)+r}\times\hfill\cr
\hfill\times\partial^r\Bigl(
\delta^{ik}J^{jl}_{2s-1}(2) - \delta^{il}J^{jk}_{2s-1}(2) +
\delta^{jk}J^{il}_{2s-1}(2) - \delta^{jl}J^{ik}_{2s-1}(2)\Bigr)\Biggr\} \cr
\noalign{\vskip5pt}
J^{ij}_{2s_1-1}\!(\!z_1\!)\!J^{kl}_{2s_2-1}\!(\!z_2\!)=\sum_{s=1}\sum_{r=0}
\Biggl\{
D_{2s-1}^{2s_1-1,2s_2-1}C_{2s-1}^{2s_1-1,2s_2-1}(r) z_{12}^{2(s-s_1-s_2)+r+1}
\times\hfill\cr
\hfill\times\partial^r\Bigl(
-\delta^{ik}J^{jl}_{2s-1}(2) + \delta^{il}J^{jk}_{2s-1}(2) +
\delta^{jk}J^{il}_{2s-1}(2) - \delta^{jl}J^{ik}_{2s-1}(2)\Bigr) \cr
+ D_{2s}^{2s_1-1,2s_2-1}C_{2s}^{2s_1-1,2s_2-1}(r) z_{12}^{2(s-s_1-s_2)
+r+2}\times\hfill\cr
\hfill\times
\partial^r(-\delta^{ik}T^{jl}_{2s}(2) + \delta^{il}T^{jk}_{2s}(2) +
\delta^{jk}T^{il}_{2s}(2) - \delta^{jl}T^{ik}_{2s}(2))\Biggr\} \cr
\hfill + N_{2s_1}^{(2)}z_{12}^{-4s_1+2}\delta_{s_1s_2}(\delta^{ik}\delta^{jl} -
\delta^{il}\delta^{jk})\hfill (2.2)\cr}
$$
where
$$C_s^{nm}(r)={(2s-1)!\, (n-m+s+r-1)!\over r!\, (2s+r-1)!\, (n-m+s-1)!}$$
and
$$D_s^{nm}={N^{(3)}_{nms}\over N^{(2)}_s}\quad .$$

What remains is to calculate the normalization constants
$N^{(2)}_s$ and $N^{(3)}_{nms}$ of the 2- and 3-point functions:
$$\eqalign{
\left\langle T^{ij}_{2s_1}(z_1)T^{kl}_{2s_2}(z_2)\right\rangle &=
N^{(2)}_{2s_1} \left(\delta^{ik}\delta^{jl} +\delta^{il}\delta^{kj}\right)
\delta_{s_1,s_2}z_{12}^{-4s_1} \cr
\left\langle J^{ij}_{2s_1-1}(z_1)J^{kl}_{2s_2-1}(z_2)\right\rangle &=
N^{(2)}_{2s_1-1} \left(\delta^{ik}\delta^{jl}
-\delta^{il}\delta^{kj}\right)\delta_{s_1,s_2} z_{12}^{-4s_1+2}\cr
\noalign{\hbox{ and say,}}
\left\langle T^{ij}_{2s_1}(z_1)T^{kl}_{2s_2}(z_2)T^{pt}_{2s_3}(z_3)
\right\rangle &= N^{(3)}_{2s_1,2s_2,2s_3}
\left(\delta^{ik}\delta^{jp}\delta^{lt}+\delta^{jk}\delta^{ip}\delta^{lt}
 +\cdots \right)\cr
&\times z_{12}^{-2(s_3-s_1-s_2)}
z_{13}^{-2(s_2-s_1-s_3)} z_{23}^{-2(s_1-s_2-s_3)}\cr}$$

In order to find $N^{(2)}$ and $N^{(3)}$ we can take the following free fermion
realization of the {\it quasiprimary currents} $T^{ij}_{2s}$ and
$J^{ij}_{2s-1}$:
$$\eqalign{
T^{ij}_{2s}(1)&=\sum_{k=1}^{2s}(-1)^k\left(\matrix{2s-1\cr k-1
\cr}\right)^2 \colon\partial^{2s-k}\psi^i(1)\partial^{k-1}\psi^j(1)\colon\cr
J^{ij}_{2s-1}(1)&= \sum_{k=1}^{2s-1}(-1)^k\left(\matrix{2s-2\cr k-1
\cr}\right)^2 \colon\partial^{2s-k-1}\psi^i(1)\partial^{k-1}\psi^j(1)\colon\cr}
\quad .\eqno(2.3)$$

The explicit form of $N^{(2)}$ and $N^{(3)}$ {\it does depend} on the concrete
construction (2.3) of  $T^{ij}_{2s}$ and $J^{ij}_{2s-1}$ we have chosen
but the ratio $D_r^{s_1s_2}$ {\it should not depend on it}. In words, the
structure
constants of the algebra are independent of the explicit realization of their
generators. It is only the central term that depends on it. Then by
straightforward computation, we get
$$\displaylines{
D_r^{s_1s_2}={(-1)^{s_1+s_2-r}\over [2(s_1+s_2-r)-2]!}\sum_{k_1=1}^{s_1}
\sum _{k_2=1}^{s_2}\sum
_{k_3=1}^{s_1+s_2-r}(-1)^{s_1+k_1+k_3}2^{s_1-s_2-k_1+k_3}\times\hfill\cr
\left(\matrix{s_1\!-\!1\cr k_1\!-\!1 \cr}\right)^2\!\! \left(
\matrix{s_2\!-\!1\cr k_2\!-\!1
\cr}\right)^2 \!\!\left(\matrix{s_1\!+\!s_2\!-\!r\!-\!1\cr k_3\!-\!1 \cr}
\right)^2\!\!(s_1\!-\!1\!-\!k_1\!+\!k_2)!\,(s_2\!-\!1\!-\!k_2\!+\!k_3)!\,
(s_1\!+\!s_2\!-\!r\!-\!1\!-\!k_3\!+\!k_1)!\; (2.4)\cr}$$

The next step is to derive the corresponding Lie algebra, encoded in the
singular terms of the OPE's (2.2). It is generated by the Laurent modes of
the currents $T^{ij}_{2s}$ and $J^{kl}_{2s-1}$:
$$\eqalign{
{\cal L}_n^{ij(2s)}&= \oint z^{2s+n-1}T^{ij}_{2s}(z)dz \quad ,\cr
{\cal Q}_n^{kl(2s-1)}&=\oint z^{2s+n-2}J^{kl}_{2s-1}(z)dz \quad .\cr}
\eqno(2.5)$$
\penalty-500
Then by simply integrating (2.2) we find the  $W_\infty (A_1)$-current algebra
in the form:
\penalty-400
$$\eqalign{
\left[ {\cal L}_{n_1}^{ij(2s_1)},{\cal L}_{n_2}^{kl(2s_2)}\right]
&=\sum_{r=0}^{ s_1+s_2-2} \Biggl\{g_{2r}^{2s_1,2s_2} (n_1,n_2)
\left(\delta\circ{\cal L}_{n_1+n_2}^{(2s_1+2s_2-2r-2)}\right)^{ijkl}\cr
%\times\hfill\cr
%\hfill\times \left( \delta^{ik}{\cal L}_{n_1+n_2}^{jl(2s_1+2s_2-2r-2)}
%+\delta^{il}{\cal L}_{n_1+n_2}^{kj(2s_1+2s_2-2r-2)} +
%\delta^{kj}{\cal L}_{n_1+n_2}^{il(2s_1+2s_2-2r-2)}+
%\delta^{jl}{\cal L}_{n_1+n_2}^{ik(2s_1+2s_2-2r-2)}\right)\cr
&+g_{2r-1}^{2s_1,2s_2}(n_1,n_2) \left(\delta\circ{\cal Q}_{n_1+n_2}^
{(2s_1+2s_2-2r-1)}\right)^{ijkl}\Biggr\}\cr
%\hfill +\delta^{il}{\cal Q}_{n_1+n_2}^{kj(2s_1+2s_2-2r+1)} +
%\delta^{kj}{\cal Q}_{n_1+n_2}^{il(2s_1+2s_2-2r+1)}+
%\delta^{jl}{\cal Q}_{n_1+n_2}^{ik(2s_1+2s_2-2r+1)}\Bigr)\Biggr\}\cr
&+N_{2s_1}^{(2)}{(2s_1+n_1-1)!\over
(4s_1-1)!(n_1-2s_1)!}\delta_{n_1+n_2}\delta_{s_1-s_2}(\delta^{ik}\delta^{jl}
 + \delta^{il}\delta^{jk})\cr
\noalign{\vskip8pt}
 \left[ {\cal L}_{n_1}^{ij(2s_1)},{\cal Q}_{n_2}^{kl(2s_2-1)}\right]
&=\sum_{r=0}^{ s_1+s_2-2} \Biggl\{g_{2r}^{2s_1,2s_2-1} (n_1,n_2)
\left(\delta\bullet{\cal L}_{n_1+n_2}^{2s_1+2s_2-2r-2)}\right)^{ijkl}\cr
%\times\hfill\cr
%\hfill\times \left( \delta^{ik}{\cal L}_{n_1+n_2}^{jl(2s_1+2s_2-2r)}
%-\delta^{il}{\cal L}_{n_1+n_2}^{kj(2s_1+2s_2-2r)} +
%\delta^{kj}{\cal L}_{n_1+n_2}^{il(2s_1+2s_2-2r)}-
%\delta^{jl}{\cal L}_{n_1+n_2}^{ik(2s_1+2s_2-2r)}\right)\cr
&+g_{2r-1}^{2s_1,2s_2-1}(n_1,n_2)\left( \delta\bullet
{\cal Q}_{n_1+n_2}^{(2s_1+2s_2-2r-3)}\right)^{ijkl}\Biggr\}\cr
%\hfill-\delta^{il}{\cal Q}_{n_1+n_2}^{kj(2s_1+2s_2-2r-1)} +
%\delta^{kj}{\cal Q}_{n_1+n_2}^{il(2s_1+2s_2-2r-1)}-
%\delta^{jl}{\cal Q}_{n_1+n_2}^{ik(2s_1+2s_2-2r-1)}\Bigr)\Biggr\}\cr
\noalign{\vskip8pt}
\left[ {\cal Q}_{n_1}^{ij(2s_1-1)},{\cal Q}_{n_2}^{kl(2s_2-1)}\right]
&=\sum_{r=0}^{ s_1+s_2-2} \Biggl\{g_{2r-1}^{2s_1-1,2s_2-1} (n_1,n_2)
\left( \delta\star{\cal Q}_{n_1+n_2}^{2s_1+2s_2-2r-3)}\right)^{ijkl}\cr
%-\delta^{ik}{\cal Q}_{n_1+n_2}^{jl(2s_1+2s_2-2r-1)}\hfill\cr
%\hfill +\delta^{il}{\cal Q}_{n_1+n_2}^{kj(2s_1+2s_2-2r-1)} +
%\delta^{kj}{\cal Q}_{n_1+n_2}^{il(2s_1+2s_2-2r-1)}-
%\delta^{jl}{\cal Q}_{n_1+n_2}^{ik(2s_1+2s_2-2r-1)}\Bigr)\cr
&+g_{2r}^{2s_1-1,2s_2-1}(n_1,n_2)\left( \delta\star {\cal L}_{n_1
+n_2}^{(2s_1+2s_2-2r-4)}\right)^{ijkl}\Biggr\}\cr
%\hfill +\delta^{il}{\cal L}_{n_1+n_2}^{kj(2s_1+2s_2-2r-1)} +
%\delta^{kj}{\cal L}_{n_1+n_2}^{il(2s_1+2s_2-2r-1)}-
%\delta^{jl}{\cal L}_{n_1+n_2}^{ik(2s_1+2s_2-2r-1)}\Bigr)\Biggr\}\cr
& +N_{2s_1-1}^{(2)}{(2s_1+n_1-2)!\over
(4s_1-3)!(n_1-2s_1+1)!}\delta_{n_1+n_2}\delta_{s_1-s_2}(\delta^{ik}\delta^{jl}
 - \delta^{il}\delta^{jk}) \cr}\eqno(2.6)
$$
where
$$
\left(\delta\circ {\cal Y}\right)^{ijkl}=\left( \delta^{ik}{\cal Y}^{jl}
+\delta^{il}{\cal Y}^{kj} +
\delta^{kj}{\cal Y}^{il}+
\delta^{jl}{\cal Y}^{ik}\right)$$

$$\left(\delta\bullet{\cal Y}\right)^{ijkl}=
\left( \delta^{ik}{\cal Y}^{jl}
-\delta^{il}{\cal Y}^{kj} +
\delta^{kj}{\cal Y}^{il}-
\delta^{jl}{\cal Y}^{ik}\right)$$

$$\left( \delta\star{\cal Y}\right)^{ijkl}=
\left( -\delta^{ik}{\cal Y}^{jl}
+\delta^{il}{\cal Y}^{kj} +
\delta^{kj}{\cal Y}^{il}-
\delta^{jl}{\cal Y}^{ik}\right)$$

The $SU(2)$ independent part of the structure constants, $g_r^{s_1,s_2}
(n_1,n_2)$ is given by:
$$
\displaylines{g_r^{s_1,s_2} (n_1,n_2)= D_r^{(s_1,s_2)}\sum_{l=0}^{r+1}
 {(-1)^l\over l!(r+1-l)!} {(2s_1-r-1+l)![2(s_1+s_2-r)-1]!
\over (2s_1-r-1)! [2(s_1+s_2-r) -1+l]!}\times\hfill\cr
\hfill \times{(s_1+n_1-1)!(s_1+s_2-r+n_1+n_2-1+l)!\over (s_1+n_1-2-r+l)!
(s_1+s_2-r+n_1+n_2-1)!}\; ,\;  (2.7) \cr}
$$
and $N_{2s}^{(2)}, N^{(2)}_{2s-1}$ are two parameters representing the central
charges of the algebra (2.6).

We should mention that in the way we arrived at the algebra (2.6) we did not
use any specific property of $SU(2)\sim O(3)$. The only specification we have
used is that the generators of the current algebra are antisymmetric tensors,
and that in their OPE's the symmetric tensors $T^{ij}_{2s}$ also contribute.
It is
evident that all $\widehat O(n)$-current algebras, for $n>2$ share these
properties.
Therefore {\it the algebra (2.6) represents the universal form of the new
family of $W_\infty$ algebras: $W_\infty (\widehat B_n)$ and $W_\infty (
\widehat D_n)$}.
Concerning the classification of the $W_\infty $-current algebras it is clear
that one has to follow the classification of the usual current algebras. For
each of them, namely $\widehat A_n^{(1)}\, ,\;\widehat A_n^{(2)}\, ,\;\widehat
 B_n,\; \widehat C_n\dots$, etc. one
has to find the OPE algebra  of all the descendents. For example, the
generators of $W_\infty (\widehat A_n)$ can be taken as having the form:
$$\eqalign{
J^A_{2s-1} &= \psi^\alpha\tau^A_{\alpha\beta} \partial^{2s-2}\psi^\beta\cr
T^{(AB)}_{2s} &=\psi^\alpha\left\{\tau^A,\tau^B\right\}_{\alpha\beta}
\partial^{2s-1}\psi ^\beta\cr }\eqno(2.8)$$
where $\tau^A_{\alpha \beta}$ are anti-hermitian matrices representing the
$SU(n)$ algebra: $\left[\tau^A,\tau^B\right] =f^{AB}_C\tau^C$. The
corresponding
$W_\infty (\widehat A_n)$ algebra have a form similar to (2.6) with the
adequate
group structure in order to conform to the $A_n$ case.  It is important to
note that
{\it the structure constants} $g_r^{s_1s_2}$ given by (2.7) {\it are universal
quantities for all the groups } $O(n), SU(n)$ etc.
\vskip 1.5cm
\penalty-400
\centerline {{\bf 3. } Symmetries of the off-critical $k=1\, , O(n) $ WZW
models}
\nobreak
\vskip.5cm
\nobreak
Our problem is to construct explicitly all the conserved charges of the models
given by (1.1), i.e. - $O(n)$-Majorana massive fermions $\overline
\psi^i(z,\overline z)\,
(i=1, \cdots, n)$. What we have to do is {\bf (1)} find the conserved
tensors $T^{ij}_{2s}$ and $J^{ij}_{2s-1}$ in terms of $\psi^i\, ,\, \overline
\psi^i$; {\bf (2)} verify that their conservation laws satisfy the criterion of
existence of noncommuting charges\ref{2}; {\bf (3)} construct these charges and
compute their algebra. One could expect that the case of $n$ massive fermions
in the $O(n)$-vector representation is a straightforward generalization of the
results for one massive fermion\ref{2}. There exist however few important
differences. The first is that together with $T_{2s}=\delta_{ij}T^{ij}_{2s}$
and $J^{ij}_{2s-1}$ we have to consider all components of the symmetric
conserved tensor $T^{ij}_{2s}$. The second very important point is that the
algebra of the standard conserved charges:
$$
\eqalign{P^{ij}_s&=\int T^{ij}_{2s}dz - \int \theta ^{ij}_{2s-2}d\overline
z\quad ,\quad \overline P^{ij}_s=\int \overline T^{ij}_{2s}d\overline z -
\int \theta ^{ij}_{2s-2}dz\cr
Q^{ij}_s&=\int J^{ij}_{2s-1}dz - \int \widetilde\theta ^{ij}_{2s-3}d\overline z
\quad ,\quad \overline Q^{ij}_s=\int \overline J^{ij}_{2s-1}d\overline z -
\int \widetilde
\theta ^{ij}_{2s-3}dz\cr}\eqno(3.1)
$$
is {\bf nonabelian}. Its abelian subalgebra is spanned by
$P_s=\delta_{ij}P^{ij}_s$ and $\overline P_s=\delta_{ij} \overline P_s^{ij}$.
In order to find this algebra, it is better to realize $P^{ij}_s\, ,\,
Q^{ij}_s$
etc in terms of differential operators. Following the standard massive fermion
technology\ref 2 we start with the ``nonquasiprimary" form of the conserved
tensors:
$$
\eqalign{T^{ij}_{2s}&=\psi^i\partial ^{2s-1}\psi^j + \psi^j\partial
^{2s-1}\psi^i\quad ,\quad T^{ij}_2={1\over 2}(\psi^i\partial \psi^j +
\psi^j\partial \psi^i)\cr
J^{ij}_{2s-1}&={1\over 2}\left(\psi^i\partial ^{2s-2}\psi^j - \psi^j\partial
^{2s-2}\psi^i \right)\cr}\eqno(3.2)
$$
and similar expressions for $\overline T^{ij}_{2s}$ and $\overline
J^{ij}_{2s-1}$. Using the equations of motion:
$$
\overline \partial \psi^k=m\overline \psi^k \quad ,\quad \partial \overline
\psi^k=-m\psi^k\eqno(3.3)
$$
one can show that (3.2) are indeed conserved tensors. For example the first few
$J^{ij}_{2s-1}$-current conservation laws are in the form:
$$
\eqalign{&\overline \partial J^{ij}_1+\partial \overline J_1^{ij}=0\cr
&\overline \partial J^{ij}_3=\partial ^2\widetilde \theta^{ij}+m^2\partial
J_1^{ij}
\cr
&\overline \partial J^{ij}_5=\partial ^4\widetilde \theta^{ij}+m^2\partial
J_3^{ij}+{3\over 2}m^2\partial ^3J_1^{ij}\cr}\eqno(3.4)
$$
etc., where $\widetilde \theta^{ij}={m\over 2}(\psi^j\overline
\psi^i-\psi^i\overline
\psi^j)$. For the ``stress-tensor's" set of conserved quantities we have:
$$
\eqalign{\overline \partial T^{ij}_2&=\partial \theta^{ij}\quad ,\quad
\theta^{ij}=m(\overline \psi^i\psi^j+\overline \psi^j\psi^i)\cr
\overline \partial T^{ij}_4&=\partial ^3\theta^{ij} + 2m^2\partial T^{ij}_2\cr
\overline \partial T^{ij}_6&=2\partial ^5\theta^{ij} + m^2\partial T^{ij}_4 +
4m^2\partial ^3T^{ij}_2\cr}\eqno(3.5)
$$
etc. Substituting (3.2) in (3.1) and using the equations of motion (3.3) one
could exclude the time derivatives in the integrands of (3.1) and then take
$t=0$ ($z=t+x, \overline z=t-x, \partial _x=\partial $). The result of these
computations is:
$$
\eqalign{P_1^{ij}&=-{1\over 2}\int dx \left[ \psi^i\partial
\psi^j+\psi^j\partial \psi^i+m(\overline \psi^i\psi^j+\overline
\psi^j\psi^i)\right]\cr
P_2^{ij}&=-{1\over 2}\int dx\left[ \psi^{(i}\partial ^3\psi^{j)} + m \overline
\psi^{(i}\partial ^2\psi^{j)} + m^2 \psi^{(i}\partial \psi^{j)} - {m^2\over 2}
\overline \psi^{(i}\partial \overline \psi^{j)} + {3\over 2} m^3 \overline \psi
^{(i}\psi^{j)}\right]\cr
Q_0^{ij}&=-\int dx \left[ \psi^{[j}\psi^{i]} + \overline \psi ^{[j}\overline
\psi^{i]}\right]\cr
Q_1^{ij}&=-\int dx\left[ 4\partial ^2\psi^{[i}\psi^{j]} - 4m\partial \overline
\psi^{[i}\psi^{j]} + m^2(\psi^{[i}\psi^{j]} + \overline \psi^{[i}\overline
\psi^{j]}\right]\cr}\eqno(3.6)
$$
etc. In order to get the momenta space form of (3.6) we have to insert the
usual creation and annihilation decomposition of $\psi^i$ and $\overline
\psi^i$:
$$
\eqalign{\psi^i(x,t)&=\int {dp\over 2\pi}\sqrt{{p_0-p\over 2p_0}}\left(
e^{ip\cdot x}a^+_i(p) + e^{-ip\cdot x}a^-_i(p)\right)\cr
\overline\psi^i(x,t)&=i\int {dp\over 2\pi}\sqrt{{p_0+p\over 2p_0}}\left(
e^{ip\cdot x}a^+_i(p) - e^{-ip\cdot x}a^-_i(p)\right)\cr}\eqno(3.7)
$$
where $ p_0^2-p^2=m^2$ and $p\cdot x =p_0 t + p x$.

The result is the following compact form of the charges (3.1):
$$
\eqalign{{\mathop{ P}^{(-\!\! -)}}_s^{ij}&={1\over 4}\int {dp\over 2\pi}(p_0\pm
p)^{2s-1}\colon (a^+_ia^-_j + a^+_ja^-_i)\colon \cr
{\mathop{ Q}^{(-\!\! -)}}_s^{ij}&=\int {dp\over 2\pi}(p_0\pm
p)^{2s}\colon (a^+_ia^-_j - a^+_ja^-_i)\colon \cr }\eqno(3.8)
$$
(we denote $p\equiv p_0 -p$ and $\overline p\equiv p_0 + p$ from now on). The
desired differential form of $P^{ij}_s$ and $Q^{ij}_s$ is a simple
consequence of (3.7), (3.8) and the anticommutation relations $\{
a^+_i(p),a^-_j(q)\} = 2\pi \delta _{ij}\delta (p-q)$ and reads:
$$
\eqalignno{\left[{\mathop{ P}^{(-\!\!-)}}^{ij}_s,\psi^k(z, \overline
z)\right]&=
(\delta^{ik}\delta^{jl}+\delta^{il}\delta^{jk}){\mathop{\partial}^{(-\!\!-)}}
^{2s-1}\psi^l(z,\overline z)&(3.9a)\cr
\left[{\mathop {Q}^{(-\!\!-)}}^{ij}_s,\psi^k(z, \overline
z)\right]&=(\delta^{ik}
\delta^{jl}-\delta^{il}\delta^{jk}){\mathop\partial^{(-\!\!-)}}
^{2s}\psi^l(z,\overline z)\quad , \quad s=1,2,\cdots &(3.9b)\cr}
$$

With eqs. (3.9) at hands we are prepared now to find the algebra of the charges
(3.1):
$$
\eqalignno{\left[Q^{ij}_{s_1},Q^{kl}_{s_2}\right]&= \delta^{ik}{ Q}^{jl}
_{s_1+s_2} +\delta^{jl}{ Q}^{ik}
_{s_1+s_2}-\delta^{il}{ Q}^{jk}
_{s_1+s_2}-\delta^{jk}{ Q}^{il}
_{s_1+s_2}&(3.10a)\cr
\left[P^{ij}_{s_1},Q^{kl}_{s_2}\right]&= \delta^{ik}P^{jl}_{s_1+s_2} - \delta
^{jl}P^{ik}_{s_1+s_2} + \delta^{il}P^{jk}_{s_1+s_2} -
\delta^{jk}P^{il}_{s_1+s_2}&(3.10b)\cr
\left[P^{ij}_{s_1},P^{kl}_{s_2}\right]&= -\delta^{ik}Q^{jl}_{s_1+s_2-1} -
\delta
^{jl}Q^{ik}_{s_1+s_2-1} - \delta^{il}Q^{jk}_{s_1+s_2-1} -
\delta^{jk}Q^{il}_{s_1+s_2-1}&(3.10c)\cr}
$$
and the same algebra for $\overline Q^{ij}_s$ and $ \overline P^{ij}_s$. The
mixed left-right algebra takes the form ($s_1\le s_2$):
$$\eqalign{
\left[P^{ij}_{s_1},\overline P^{kl}_{s_2}\right]&= -(-m^2)^{2s_1-1}(\delta^{ik}
\overline Q^{jl}_{s_2-s_1} + \delta
^{jl}\overline Q^{ik}_{s_2-s_1} +\delta^{il}\overline Q^{jk}_{s_2-s_1} +
\delta^{jk}\overline Q^{il}_{s_2-s_1})\cr
\left[P^{ij}_{s_1},\overline Q^{kl}_{s_2}\right]&= (-m^2)^{2s_1-1}(\delta^{ik}
\overline P^{jl}_{s_2-s_1+1} - \delta
^{jl}\overline P^{ik}_{s_2-s_1+1} + \delta^{il}\overline P^{jk}_{s_2-s_1+1} -
\delta^{jk}\overline P^{il}_{s_2-s_1+1})\cr
\left[Q^{kl}_{s_1},\overline P^{ij}_{s_2}\right]&= -(-m^2)^{2s_1}(\delta^{ik}
\overline P^{jl}_{s_2-s_1} - \delta
^{jl}\overline P^{ik}_{s_2-s_1} + \delta^{il}\overline P^{jk}_{s_2-s_1} -
\delta^{jk}\overline P^{il}_{s_2-s_1})\cr
\left[Q^{ij}_{s_1},\overline Q^{kl}_{s_2}\right]&= (-m^2)^{2s_1}(\delta^{ik}
\overline Q^{jl}_{s_2-s_1} + \delta
^{jl}\overline Q^{ik}_{s_2-s_1} - \delta^{il}\overline Q^{jk}_{s_2-s_1} -
\delta^{jk}\overline Q^{il}_{s_2-s_1})\cr}\eqno(3.11a,b,c,d)
$$

Using the formal identity $\overline \partial =-m^2\partial ^{-1}$ one can
write $(m^2)^{-s}Q_s^{ij}\equiv \widetilde Q_s^{ij}$ and $(m^2)^{-s}\overline
Q^{ij}_s\penalty-200\equiv\widetilde Q_{-s}^{ij}$ as a unique object
$\widetilde Q_s^{ij}\,
(-\infty \le s \le \infty)$ which
{\it generates the $O(n)$-Kac-Moody algebra (3.10a), (3.11d)}. One can
recognize
the total algebra (3.10),  (3.11) as a {\it subalgebra $GL(n,R)_{mod\, 2}$
of the $\widehat {GL}(n,R)$-Kac-Moody algebra } spanned
by $\widetilde P^{ij}_{2s-1} \equiv P^{ij}_s$
and $\widetilde Q^{ij}_{2s}=Q^{ij}_s$, i.e. the {\it closed subalgebra of
symmetric
generators $P^{ij}$ with odd indices and antisymmetric generators $Q^{ij}$ with
even indices}.

One can derive the algebra (3.10), (3.11) using directly (3.8). The advantage
is that in this way we calculate the value of the central charge of the $
\widehat
O(n)$-Kac-Moody algebra. Starting from (3.8) and taking care about the right
normal ordering in the r.h.s. of the commutator $[Q^{ij}_s,\overline Q^{ij}_s]$
 we find the central term in the form:
$$(-m^2)^{2s} {s\over 2} (\delta^{ik}\delta^{jl}-\delta^{il} \delta^{jk})
\quad .$$

The central charge therefore is $p=1$, i.e. the same as for the massless $O(n)$
-fermions. The appearance of non-zero central charge has far-reaching
consequences
in the application of the degenerate higher weight representations of (3.10),
(3.11) for constructing the exact solutions of these models.

Having constructed the algebra of the conserved charges (3.1) we have to answer
the question about its meaning. As in the case of one Majorana fermion\ref{2},
one can expect that the {\it charges
$\oversome{(-\!\!-)}\! P^{\,\,ij}_s$ and $\oversome{(-\!\!-)}\! Q_s^{\,\,ij}$
are generators of specific new symmetries of the $O(n)$-massive fermions action
(1.1)}. It is indeed the case and the proof of this statement  goes trough
simple higher derivatives
computations. We shall present here only one detail of this proof. Consider the
commutator:
$$
\left[Q^{kl}_s,\overline
\psi^j\psi^j\right]=(\delta^{kj}\delta^{lm}-\delta^{km}\delta^{lj})\left[
(\partial^{2s}\overline \psi^m)\psi^j - \overline \psi^m\partial
^{2s}\psi^j\right]\eqno(3.12)$$
The conclusion that the r.h.s. of (3.12) is a total derivative is based on
the following identity:
$$
(\partial ^{2s}\overline \psi^m)\psi^j - \overline \psi^m\partial
^{2s}\psi^j=\partial \left\{ \sum^{s-1}_{l=0}(-1)^l[\partial ^{2s-l-1}\overline
\psi^m\partial ^l\psi^j - \partial ^l\overline \psi^m\partial
^{2s-l-1}\psi^j]\right\}\quad .
$$
 The same is true
for the commutators $[\overline Q^{kl}_s,\overline \psi^j\psi^j], [\oversome
{(-\!\!-)}\!Q^{\,\,kl}_s,\psi^j\overline \partial \psi^j]$, and
$ [\oversome{(-\!\!-)}\!
Q^{\,\,kl}_s,\overline \psi^j\partial \overline \psi^j]$ as well.
Therefore the {\it action (1.1) is invariant under  the infinitesimal $\widehat
O(n)$-Kac-Moody transformation (3.9b)}:
$$\left[\widetilde Q^{ij}_s, S\right]=0\quad -\infty \le s \le \infty \quad .$$
It is straightforward to make an analogous conclusion concerning the
$P_s^{ij}$-symmetries of (1.1). Then we have to complete our statement about
the symmetries of (1.1): the {\it subalgebra (3.10), (3.11) of $\widehat
{GL}(n,R)$-Kac-Moody algebra appears as a larger algebra of symmetries of the
$O(n)$-massive Majorana fermions action (1.1)}.

The question about the origin of such symmetries is in order. The first to be
noted is that they are not related to the local $GL(n,R)$-gauge
transformations in two dimensional $(z,\overline z)$ space. However the
momenta space form of the transformations (3.9):
$$\eqalign{
[\widetilde P^{ij}_s, a^\pm_k(p)] &=\pm i(\delta^{ik}\delta^{jl}+\delta^{il}
\delta^{jk})p^{2s-1}a^\pm_l(p)\cr
[\widetilde Q^{ij}_s, a^\pm_k(p)] &=\pm i(\delta^{ik}\delta^{jl}-\delta^{il}
\delta^{jk})p^{2s}a^\pm_l(p)\quad ,\quad -\infty \le s \le \infty \cr}
\eqno(3.13)
$$
is very suggestive. Remember the standard realization (1.5) of the $x$-space
infinitesimal local gauge transformations. Therefore the specific
$\widehat{GL}(n,R)_{mod\, 2}$-Kac-moody algebra of
symmetries of (1.1) we have found
a manifest  local momenta-space $GL(n,R)$ gauge invariance. The parameters
$\omega^{ij}(p)$ of these transformations are restricted under the conditions:
$$
\eqalign{
\omega^{(ij)}(p)=-\omega^{(ij)}(-p)\cr
\omega^{[ij]}(p)=\omega^{[ij]}(-p)\quad .\cr}
$$
These properties are not a specific feature of the $O(n)$-massive fermions
only. One can find {\it two incomplete} $(s\ge 0!)\, GL(n,R)_{mod\, 2}$
local  (in momenta
space) gauge groups of symmetries for the massless $O(n)$-fermions as well. One
can further speculate that the conformal $W_\infty (\widehat G_n)$-algebra
describes the symmetries of the phase space of the corresponding conformal
model. A part of these symmetries survives the perturbation forming the {\it
nonconformal $\widetilde W_\infty (\widehat G_n)$-algebra}. The $x$-space
symmetries
are restricted now to $2-D$ Poincar\'e group, global $G_n$ gauge invariance and
the specific $\widetilde W_\infty(\widehat G_n)$ symmetries, for example the
$GL(n,R)_{{\rm mod} 2}$-Kac-Moody algebra (3.10), (3.11). The latter manifests
as a local
gauge transformations in the $p$-space. In this line of arguments one can
consider $\widehat{GL} (n,R)_{{\rm mod} 2}$
(and $\widetilde W_\infty(G_n)$ in general) as symmetries
of the phase space of the integrable models.

Our motivation to study the full set of conservation charges for $k=1$,
$O(n)$-WZW massive models was to find the full algebra of the symmetries of
the model
({\it presumably noncommuting}) in order to use it for the calculations of the
correlation functions. We already have found one nontrivial subalgebra (3.10),
(3.11), generated by the charges of the conserved tensors
$\oversome{(-\!\!-)}\!
T^{\,\,ij}_{2s}$ and $\oversome{(-\!\!-)}\! J^{\,\,ij}_{2s-1}$. Is this
symmetry
sufficient to fix all the correlation functions (without using equation of
motion)? What is known from the conformal WZW models is that the conformal
current algebra ${\rm Vir} \subset\!\!\!\!\!\!\times
 \widehat G_n$ is a powerful tool for such calculations. Therefore we have to
look for more {\it new charges} $L_n$, generating the  Virasoro algebra, in
order to complete our $\widehat O(n)$ (or $\widehat{GL}(n,R)_{{\rm mod} 2}$)
-Kac-Moody algebras
(3.10), (3.11) to the  larger ${\rm Vir} \subset\!\!\!\!\!\!\times
 \widehat O(n)$ algebra.

How to construct the Virasoro charges for one massive
fermion we already know from the
off-critical Ising model case. In order to generalize it for the
$O(n)$-massive fermions we have to find specific combinations of the ``higher
momenta" of the $\oversome{(-\!\!-)}\! T^{\,\,ij}_{2s},\oversome{(-\!\!-)}\!
J^{\,\,ij}_{2s-1},\theta^{ij}$ and $\widetilde \theta^{ij}$ to be conserved.
{}From the
explicit form (3.4) and (3.5) of these standard conservation laws one can
conclude that they satisfy the criterion\ref{2} for existence of
new charges.  Therefore we can construct
$(4s-3){n(n+1)\over 2}$ new symmetric charges  $L^{ij(2s)}_{-n},
\overline L^{ij(2s)}_{-n}\, (0\le n\le 2s-1)$ and
$(4s-5){n(n-1)\over 2}$ antisymmetric ones  $Q^{ij(2s-1)}_{-k}, \overline
Q^{ij(2s-1)}_{-k}\, (0\le k\le 2s-2)$ for each
$s=2,3, \cdots$.

To begin with the first momenta, i.e. quantities linear in $z$ and $\overline
z$. The simplest one is the generalization  of the Lorentz rotation
$L_0={1\over 2}\delta^{ij}L_0^{ij}$:
$$
L_0^{ij}=\int (zT^{ij}_2 + \overline z\theta^{ij})dz - \int (\overline z
\overline T^{ij}_2 + z\theta^{ij})d\overline z\quad .
$$
The next two are the off-critical analogs of the conformal first momenta of
$T^{ij}_4$ and $\overline T^{ij}_4$:
$$
\eqalign{L^{ij(4)}_{-2}&=\int (zT^{ij}_4 + 2m^2\overline zT^{ij}_2)dz - m^2\int
(2zT^{ij}_2 + \overline z\theta^{ij})d\overline z\cr
\overline L^{ij(4)}_{-2}&=\int (\overline z\overline T^{ij}_4 + 2m^2 z\overline
T^{ij}_2)d\overline z - m^2\int
(2\overline z\overline T^{ij}_2 + z\theta^{ij})dz\cr}\quad .
$$

One can go further and construct
$\oversome{(-\!\!-)}\,L^{\,\,ij(6)}_{-4}$ etc.
However all of them are straightforward $O(n)$-matrix generalization of the
corresponding one fermion charges
$\oversome{(-\!\!-)}\, L^{\,\,(2s)}_{-2s+2}={1\over 2}\delta^{ij}
\oversome{(-\!\!-)}\, L^{\,\,ij(2s)}_{-2s+2}$
and we can take them in the following differential form:
$$
\eqalign{\left[L^{ij(2s)}_{-2s+2},\psi^k(z,\overline z)\right]&=
-i(\delta^{ik}\delta^{jl}
+ \delta^{il}\delta^{jk})\left( \overline z\overline \partial -z\partial
-{2s-1\over 2} \right)\partial ^{2s-2}\psi^l\cr
\left[\overline L^{ij(2s)}_{-2s+2},\psi^k(z,\overline z)\right]&=
-i(\delta^{ik}\delta^{jl} +
\delta^{il}\delta^{jk})\left(\overline z\overline \partial -z\partial
+{2s-3\over 2} \right)\overline\partial^{2s-2} \psi^l\cr  }\eqno(3.14)
$$
The proof that they are indeed the conserved charges we are looking for is
again based on the fact that they do generate new symmetries of the action
(1.1), i.e.
$$[L^{ij(2s)}_{-2s+2},S]=0=[\overline L^{ij(2s)}_{-2s+2},S]
$$
The last statement follows from specific higher derivatives identities similar
to the one used in the proof of (3.12)

The question about the algebra of these new symmetries is now in order.  By
direct calculations, using (3.14) one can see
that $L^{ij(2s)}_{-2s+2}$ and $ \overline L^{ij(2s)}_{-2s+2}$ {\it does not
close} an algebra. It is necessary to consider together with them the first
momenta
$\oversome{(-\!\!-)}\,{Q}^{\,\,ij(2s-1)}_{-2s+3}$ of the current
$J^{ij}_{2s-1}$. Before doing this we should mention that the traces
$\oversome
{(-\!\!-)}\, L^{\,\,(2s)}_{-2s+2}={1\over 2}\delta_{ij}\oversome{(-\!\!-)}\,
L^{\,\,ij(2s)}_{-2s+2}
$ {\it do close} an algebra which coincides with the off-critical Virasoro
algebra $V_c$  of the off-critical Ising model\ref{2}. One could
wonder what is then the algebra of $\widetilde Q^{ij}_s$ and  these Virasoro
generators:
$$
V_k={1\over 4}(-m^2)^k\delta_{ij}L^{ij(2k+2)}_{-2k}\quad ,\quad
V_{-k}={1\over 4}(-m^2)^k\delta_{ij}\overline L^{ij(2k+2)}_{-2k}$$

As one could expect the result of simple computations is the larger current
algebra $V_c\subset\!\!\!\!\!\!\times \widehat O_n $:
$$
\eqalign{
\quad\Bigl[V_{m_1},V_{m_2}\Bigr] =& (m_1-m_2)V_{m_1+m_2}+{n\over
24}m_1(m_1^2-1)\delta_{m_1+m_2}\cr
\quad\Bigl[V_{m_1},\widetilde Q_{m_2}^{ij}\Bigr] =&
-m_2\widetilde Q^{ij}_{m_1+m_2}\cr
\quad\Bigl[\widetilde Q_{m_1}^{ij},\widetilde Q_{m_2}^{kl}\Bigr] =&\delta^{ik}
\widetilde
Q^{jl}_{m_1+m_2} + \delta^{jl}\widetilde Q^{ik}_{m_1+m_2} -
\delta^{il}\widetilde
Q^{jk}_{m_1+m_2} - \delta^{jk}\widetilde Q^{il}_{m_1+m_2} +\cr
&+ {n\over 2} m_1
\delta_{m_1+m_2}(\delta^{ik}\delta^{jl} - \delta^{il}\delta^{jk})\quad .\cr}
\eqno(3.15) $$
We have enlarged  in this way the known symmetries of the action (1.1) to the
$V_c\subset\!\!\!\!\!\!\times  \widehat O(n)$-algebra.

Turning back to our problem of constructing the first momenta of the current
 $J^{ij}_{2s-1}$ we start with the explicit form of the simplest two of them:
$$
\eqalign{Q^{ij(3)}_{-1}&= \int (zJ_3^{ij}-2m^2\overline zJ_1^{ij})dz - \int
\left[ 2\widetilde \theta ^{ij}-2m^2zJ_1^{ij}-\partial (z\widetilde
\theta^{ij})+2m^2z\overline J_1^{ij}\right] d\overline z\cr
Q^{ij(5)}_{-3}&=\int(zJ_5^{ij} + m^2 \overline zJ_3^{ij})dz - \int
[m^4\overline zJ_1^{ij} - {9\over 2}m^2 \partial J_1^{ij} + {3\over 2} m^2
\partial ^2 (z J_1^{ij}) + m^2 z J_3^{ij} \cr
&+ m^2\partial (\overline z\widetilde \theta
^{ij})+\partial^3(z\widetilde\theta^{ij})
- 4 \partial ^2 \widetilde \theta^{ij}] d\overline z\cr}
$$
Following the method we used above for $\widetilde P_s^{ij}$ and $\widetilde
Q_s^{ij}$ we arrive at the following general differential form for
$\oversome{(-\!\!-)}\,Q^{\,\,ij(2s-1)}_{-2s+3}$:
$$
\eqalign{
\left[Q^{ij(2s-1)}_{-2s+3},\psi^k(z, \overline z)\right]&=-i(\delta^{ik}\delta^
{jl} -\delta^{il}\delta^{jk})(\overline z\overline \partial - z \partial
-s+1)\partial ^{2s-3}\psi^l(z,\overline z)\cr
\left[\overline Q^{ij(2s-1)}_{-2s+3},\psi^k(z,\overline z)\right]&=-i(\delta^
{ik}\delta^{jl} -\delta^{il}\delta^{jk})(\overline z\overline \partial -
z \partial + s-2)\overline \partial ^{2s-3}\psi^l(z,\overline z)\cr}\eqno(3.16)
$$

Considering $\oversome{(-\!\!-)}\,Q^{\,\,ij(2s-1)}_{-2s+3}\!$
together with
$\oversome {(-\!\!-)}\,L^{\,\,ij(2s)}_{-2s+2}$,
$\oversome{(-\!\!-)}\,Q^{\,\,ij}_{s}\!
\equiv\!\oversome{(-\!\!-)}\,Q^{\,\,ij(2s-1)}_{-2s+2} \,
\oversome{(-\!\!-)}\,P^{\,\,ij}_{s}\!\equiv\! \oversome
{(-\!\!-)}\,L^{\,\,ij(2s)}_{-2s+1}$
we are expecting them to close an algebra. However this is not the case. One
can easily check using (3.16) that the commutator
$[Q^{ij(2s_1-1)}_{-2s_1+3},Q^{kl(2s_2-1)}_{-2s_2+3}]$ contains higher momenta
of $J^{ij}_{2s-1}$ and $T^{ij}_{2s} $ as well. For example, the simplest one
has the form:
$$\eqalign{
\left[Q_{-1}^{ij(3)},Q_{-1}^{kl(3)}\right]&=-\delta^{ik}(Q^{jl(5)}_{-2}-
4Q^{jl(3)}_{-2})+\delta^{il}(Q^{jk(5)}_{-2}-4Q^{jk(3)}_{-2})\cr
& + \delta^{jk}(Q^{il(5)}_{-2}-4Q^{il(3)}_{-2}) - \delta^{jl}(Q^{ik(5)}_{-2}
-4Q^{ik(3)}_{-2})\cr}
\quad .\eqno(3.17)
$$
It includes together with the ``zero momenta" $Q^{il(3)}_{-2}\equiv Q^{il}_1$,
the {\it second momenta} $Q^{ij(5)}_{-2}$ of $J^{ij}_5$:
$$\eqalign{
\left[ Q^{ij(5)}_{-2},\psi^k(z, \overline z)\right]
&=-i(\delta^{ik}\delta^{jl}-\delta^{il}\delta^{jk})\left[ {15\over 4} +
(\overline z\overline \partial - z\partial -3/2)^2\right] \partial ^2\psi^l\cr
&=-{i\over 2}(\delta^{ik}\delta^{jl}-\delta^{il}\delta^{jk})\left[
(\overline z\overline \partial - z\partial)_2 + (\overline z\overline \partial
-z\partial -4)_2 \right]\partial ^2\psi^l\cr}\quad ,\eqno(3.18)
$$
where $(A)_n=A(A+1) \cdots (A+n-1)$.

All this discussion is to demonstrate that {\it the algebra of the first
momenta} of $T^{ij}_{2s}$ and $J^{ij}_{2s-1}$ {\it is not closed}. Involving
the higher momenta of $T^{ij}_{2s}\, ,\, J^{ij}_{2s-1}$ we are constructing in
this way an algebra of the $W_\infty (\widehat G_n)$-type (2.6). Leaving the
problem of the general structure of the algebra of all the symmetries of (1.1)
to the next section we address here the question about its subalgebras. Up to
now we have constructed two such subalgebras: $\widehat{GL}(n,R)_{{\rm mod} 2}$
given by (3.10), (3.11) and ${\rm Vir} \subset\!\!\!\!\!\!\times\widehat O(n)$
of
eq. (3.15). Deriving the missing commutator
$$
\left[ V_{m_1},\widetilde P^{ij}_{m_2}\right] = -(m_2-1/2)\widetilde
P^{ij}_{m_1+m_2}\eqno(3.19)
$$
we can unify them in an unique current algebra, namely: ${\rm Vir}
\subset\!\!\!\!\!\!\times \widehat {GL}(n,R)_{{\rm mod} 2}$. Are there more
subalgebras of this type? As in the cases of one\ref{2} and two\ref{1}
fermions one could expect to find two incomplete $(n\ge -1)$ Virasoro
subalgebras. In our case they are generated by a specific combination of
$\delta^{ij} \oversome{(-\!\!-)}\,L^{\,\,ij(2k)}_{-s+1}$ and $ \delta^{ij}
\oversome{(-\!\!-)}\,L^{\,\,ij}_{-1}\equiv\oversome{(-\!\!-)}\,P_1$:
$$
{\mathop{\cal L}^{(-\!\!-)}}_n =\sum ^n_{k=\left[{n\over 2}\right]} \beta_k
{\mathop{L}^{(-\!\!-)}}^{ij(2k)}_{-n+1}\delta_{ij}=\left[ \overline z\overline
\partial -z\partial \pm 1/2\right] _{n+1}{\mathop{\partial}^{(-\!\!-)}} ^n
\quad .\eqno(3.20)
$$
Do they have an $\widehat O(n)$-Kac-Moody counterpart? The form of the
commutator
(3.17) suggests to consider $Q_0^{ij}\, ,\, Q^{ij(3)}_{-1}$ and $Q^{ij(5)}_{-2}
- 4Q^{ij(3)}_{-2}$ as  appropriate  candidates for the generators ${\cal Q}^
{ij}_0\, ,\, {\cal Q}^{ij}_1$ and $ {\cal Q}^{ij}_2$ of the  incomplete $(n\ge
0)$ $\widehat O(n)$-Kac-Moody algebra. The differential form of these
generators
$$
\eqalign{
({\cal Q}^{ij}_0)_{kl}&=-i(\delta^{ik}\delta^{jl} -\delta^{il}\delta^{jk})\cr
({\cal Q}^{ij}_1)_{kl}&=({\cal Q}^{ij}_0)_{kl}(\overline z \overline \partial
-z\partial -1)\partial\cr
({\cal Q}^{ij}_2)_{kl}&=({\cal Q}^{ij}_0)_{kl}(\overline z \overline \partial
-z\partial -1)(\overline z \overline \partial - z\partial -2)\partial^2\cr}
$$
allows us to guess the general form of the $\widehat O(n)$ generators:
$$
({\cal Q}^{ij}_s)_{kl}=({\cal Q}^{ij}_0)_{kl}
\left[ \overline z \overline \partial - z\partial -1\right] _n\partial^n \quad
,\quad n\ge 0\quad .\eqno(3.21)
$$
Using the simple identity:
$$
\partial ^k\left[ \overline z\overline \partial -z\partial -1\right] _n=\left[
\overline z \overline \partial  - z\partial -1-k\right] _n \partial ^k
$$
one can easily verify that (3.21) indeed close $\widehat O(n)$-Kac-Moody
algebra
(3.10a). Similarly, the conserved charges
$$
(\overline {\cal Q}^{ij}_s)_{kl}=({\cal Q}^{ij}_0)_{kl}
\left[ \overline z \overline \partial - z\partial \right] _n\overline
\partial^n \quad ,\quad n\ge 0
$$
generate one more $\widehat O(n)$-current algebra. These two algebras however
do not
mutually commute.

There are certain indications that the algebras of symmetries of (1.1) we have
described up to now are sufficient for the calculation of the correlation
functions. Leaving aside the problem of how to use the null-vector
corresponding to
(3.15), (3.19) or how to solve the infinite system  of Ward identities (W.I.'s)
for the charges $\widetilde P^{ij}_s\, ,\, \widetilde Q^{ij}_s\, ,\, V_s\, ,\,
{\cal L}_n $ and $\oversome{(-\!\! -)}\,{\cal Q}^{\,\,ij}_n$ we shall mention
the
following simple and {\it remarkable fact}: the $Q^{ij(3)}_{-1}$ (or $\overline
Q^{ij(3)}_{-1}$) W.I.'s for the 2-point function
$$
g^{lm}(z_1,z_2\vert\overline z_1,\overline
z_2)=\left\langle\psi^l(z_1,\overline z_1) \psi^m(z_2,\overline
z_2)\right\rangle
$$
coincide with the $K_1$-Bessel equation.

Taking into account the Poincar\'e invariance (i.e., $L_0$, $\oversome
{(-\!\!-)}\, L_{-1}={1\over 2}\delta^{ij}\oversome{(-\!\!
-)}\,L_{-1}^{\,\,ij}$)
 we get
$$
g^{lm}(z_1,z_2\vert\overline z_1,\overline z_2)=\delta^{lm}\sqrt{\overline
z_{12}\over z_{12}} K(x)\quad ,\quad x=\sqrt{-4m^2z_{12}\overline z_{12}}
$$
We next require the  $Q_{-1}^{ij(3)}$-Ward identity:
$$\left\langle Q_{-1}^{ij(3)}\psi^l(z_1,\overline z_1)\psi^m(z_2,\overline
z_2)\right\rangle =0 \quad .
$$
As a consequence of (3.16) and $Q^{ij(3)}_{-1}$-invariance of the vacua we
obtain the following equation:
$$
\left(\overline z_{12}+{2\over m^2}\partial_{12}+{1\over
m^2}z_{12}\partial_{12}^2\right)\sqrt{ \overline
z_{12}\over z_{12}} K(x)=0
$$
which is equivalent to the  $K_1$-Bessel equation:
$$
x^2K^{\prime\prime}(x)+xK^\prime (x)-(x^2+1)K(x)=0
$$
We have in fact to solve an infinite system of higher order differential
equations representing the remaining Ward identities:
$$
\left\langle O_n\psi^l(z_1,\overline z_1)\psi^m(z_2,\overline
z_2)\right\rangle =0
$$
where $O_n=\{ \widetilde P^{ij}_s,\widetilde Q^{ij}_s,V_s,{\cal L}_n,
\oversome {(-\!\!-)}\, {\cal Q}^{\,\,ij}_n\}$.
The algebraic explanation of why all they could
be solved in terms of $K_1$ is not known. One  non-trivial check is the
$L_{-2}^{ij(4)}$ and ${\cal L}_1$-W.I.'s which are specific
 third order differential equation. In the $L^{ij(4)}_{-2}$ case for example
we have:
$$
x^3K^{\prime\prime\prime}(x)+2x^2K^{\prime\prime}(x)-x(x^2+1)K^\prime (x)
-(x^2+1)K(x) =0
$$
As it has been demonstrated in [2], this equation has $K_1(x)$ as a solution.
\penalty-500
\vskip 1cm
\penalty-500
\centerline {{\bf 4.} Off-critical $\widetilde W_\infty(\widehat G_n)$-algebra}
\nobreak
\vskip .5cm
\nobreak
Although it seems reasonable that the ${\rm Vir}
\subset\!\!\!\!\!\!\times\widehat{GL} (n,R)_{{\rm mod} 2}$ and the other $
\widehat
O(n)$-Kac-Moody and Virasoro algebras)  are the most important part of the
symmetries of (1.1) we find interesting to study the full algebra as well. To
do
this we have to continue with the constructions of the higher momenta of
$T^{ij}_{2s}$ and $J^{ij}_{2s-1}$. One such example is the second momenta of
$J^{ij}_3$:
$$
Q_0^{ij(3)}=\int {\cal F}_0^{ij} dz -\int {\cal G}_0^{ij} d\overline z$$
where
$$\eqalign{{\cal F}_0^{ij} &=z^2J^{ij}_3 -4m^2z\overline z
J_1^{ij}+2m^2z^2\overline J_1^{ij} -\overline\partial\left( \overline
z^2\widetilde\theta^{ij}\right) +4\overline z\widetilde\theta^{ij}\cr
{\cal G}_0^{ij} &= -\overline z^2\overline J_3^{ij}+4m^2z\overline z\overline
J_1^{ij}-2m^2z^2J_1^{ij}-\partial\left( z^2\widetilde\theta^{ij}\right)
+4z\widetilde\theta^{ij}\cr}
$$
are components of conserved tensor:
$$
\overline\partial{\cal F}_0^{ij}=\partial {\cal G}_0^{ij}
$$
One can realize $Q^{ij(3)}_0$ as differential operator as well:
$$
({\cal Q}^{ij(3)}_0)_{kl}=-{1\over 2}({\cal Q}^{ij}_0)_{kl}
\left[ (\overline z \overline \partial - z\partial )_2 +
(\overline z \overline \partial - z\partial -2)_2\right] \quad .
$$

This formula together with (3.16) and (3.18) allows us to make a conjecture
about the general form of all the conserved charges related to $J^{ij}_{2s-1}$
$$
\displaylines{
\quad\left( { Q}^{ij(2s-1)}_{-m}\right)_{kl}
=-{1\over 2}\left( { Q}^{ij}_0
\right)_{kl}\Bigl[ (\overline z \overline \partial - z\partial + \alpha
)_{2s-2-m} + \hfill\cr
\hfill + (\overline z \overline \partial - z\partial +\alpha -2s + 2)_{2s
-2-m}\Bigr] \partial ^m \; \phantom{(4.2)}\cr
\quad\left( \overline { Q}^{ij(2s-1)}_{-m}\right)_{kl}
=-{1\over 2}\left(
{ Q}^{ij}_0\right)_{kl}\Bigl[ (\overline z \overline \partial -
z\partial
+\overline \alpha -2s+m+3)_{2s-2-m} +\hfill\cr
\hfill  + (\overline z
\overline \partial -
z\partial\! +\!\overline \alpha \!+\! m\!+\!1)_{2s-2-m}\Bigr]
\overline\partial ^m \; \phantom{(4.2)}\cr
\phantom{\quad\left( {\cal Q}^{ij(2s-1)}_{-m}\right)_{kl} }
0\le m\le 2s-2\hfill (4.2)\cr}
$$
where $\alpha =0\, ,\, \overline \alpha =-1$ for $\psi$ and $\alpha =1\, ,\,
\overline \alpha=0$ for $\overline \psi$. The proof of this conjecture is again
indirect. By tedious higher derivatives calculus and identities similar to the
one used before (for the case $Q^{ij}_s$) one can verify that (4.2) are indeed
symmetries of (1.1), i.e. $\left[ \oversome {(\!\!-)}\,Q^{\,\,ij(2s+1)}_{-m},
S\right] =0$.

We have next to find the differential form of the remaining part
$ \oversome{(\!\!-)}\,L_{-m}^{\,\,ij(2s)}\, (0\le m\le 2s-1)$
of the  generators. We
have already mentioned  their close relation with the one fermion's generators
$ \oversome{(\!\!-)}\,L_{-m}^{\,\,(2s)}$ (see ref.[2]).
This fact allows us to write
$ \oversome{(\!\!-)}\,L_{-m}^{\,\,ij(2s)}$ using the
$ \oversome{(\!\!-)}\,L_{-k}^{\,\,(2s)}$ differential operators:
$$
\eqalign{
\left( L_{-m}^{ij(2s)}\right)_{kl} &= -{1\over 2}\!\left(
I^{ij}\right)_{\!kl} \!
\left[ \left(\overline z\overline \partial-z\partial  +\alpha\right)_{2s-1-m} +
\left(\overline z\overline \partial -z\partial + \alpha -2s+1\right)_{2s-1-m}
\right]\partial^m\cr
\left( \overline L_{-m}^{ij(2s)}\right)_{kl} &= -{1\over 2}\!
\left( I^{ij}\right)
_{\!kl}\!\left[ \!\!\left(\overline z\overline \partial\!-\!z\partial \! +\!
\overline\alpha\! -\!
2s\! + \! m \!+ \! 2 \right)_{2s-1-m}\!\! +\!
\left(\overline z\overline \partial \!-\! z\partial \!+\! \overline\alpha \! +
\! m\! +\! 1\right)_{2s-1-m} \right]\overline \partial^m\cr}\eqno(4.3)
$$
where $\left(I^{ij}\right)_{kl}=\delta^{ik}\delta^{jl}+\delta^{il}\delta^{jk}$.

We have exhausted in this way all the symmetries of the $O(n)$-massive free
fermions. What we are going to show now is that {\it their algebra consists of
two noncommuting (incomplete $0\le k\le 2s-1$)} $W_\infty(\widehat G_n)${\it
algebras}. In words the conformal structure constants (2.7) reappear again in
the massive theory as structure constants of the ``left" and ``right"
subalgebras spanned by $L^{ij(2s)}_{-m},Q^{ij(2s-1)}_{-m}$ and $\overline
L^{ij(2s)}_{-m},\overline Q^{ij(2s-1)}_{-m}$. The method is similar to the one
used in the case of $W_\infty(V)$-algebra\ref{2}. We start with the ``conformal
decomposition" of the ``left" generators:
$$
\eqalign{
L_{-k}^{ij(2s)} &=\sum_{l=0}^{2s-1-k}\left(\matrix{ 2s-1-k\cr l\cr}
\right) \left(\overline z\overline\partial^2 +\alpha\overline\partial\right)^l
{\cal L}_{-k-l}^{ij(2s)} (-m^2)^{l}\, ,\cr
Q_{-k}^{ij(2s-1)} &=\sum_{l=0}^{2s-2-k}\left(\matrix{ 2s-2-k\cr l\cr}
\right) \left(\overline z\overline\partial^2 +\alpha\overline\partial\right)^l
\widetilde {\cal Q}_{-k-l}^{ij(2s-1)} (-m^2)^{l}\, .\cr }\eqno(4.4)
$$
In the calculation of the commutators of (4.4) we are using the conformal
$CR$'s
(2.6) and the fact that the operators $(\overline z\overline \partial ^2 +
\alpha \overline \partial )^l$ are commuting. In order to prove that
$L^{ij(2s)}_{-m}$ and $Q^{ij(2s-1)}_{-m}$ satisfy (2.6) with the same structure
constants we do need the following property of $g_r^{s_1s_2}(n_1,n_2)$ to be
satisfied:
$$\displaylines{
g_r^{s_1,s_2} (-k_1,-k_2)\!\!\left(\matrix{s_1+s_2-r-k_1-k_2
-1\cr
n\cr}\right)=\hfill\cr
\hfill=\sum_l\left(\matrix{s_1-k_1-1\cr l\cr}\right)
\left(\matrix{s_2-k_2-1\cr n-l\cr}\right) g_r^{s_1,s_2}(-k_1-l,-k_2-n+l)\quad
(4.5)\cr}$$

We have to remember at this point that the only structure constants we know are
the ones calculated in the quasiprimary basis (2.3). Therefore we are forced to
use them in the proof of (4.5). The question now is whether the identity (4.5)
depends on the basis. The answer is that (4.5) holds in all the basis. It is
related to the fact that the form of the conformal decomposition (4.4) is
universal. To see it we have to write all the generators (conformal and
nonconformal) in the quasiprimary basis.
$$
\eqalign{
\left( \widetilde {\cal L}_{-m}^{ij(2s)}\right) _{kl} &=\left(
I^{ij}\right)_{kl} \sum ^{2s}_{p=1} (-1)^p \left( \matrix{2s-1\cr
p-1\cr}\right)^2 \partial ^{p-1}(z^{2s-m-1}\partial ^{2s-p})\cr
\left( \widetilde L_{-m}^{ij(2s)}\right) _{kl} &=\left(
I^{ij}\right)_{kl} \sum ^{2s}_{p=1} (-1)^p \left( \matrix{2s-1\cr
p-1\cr}\right)^2 (\overline z\overline \partial -z\partial -p+1)_{2s-m-1}
\partial ^m \cr}\eqno(4.6)
$$
etc. The crucial point is that using (4.6) we arrive again at the same
``conformal decomposition" (4.4). The last step of the proof is to substitute
(2.7) in (4.5)  and verify that it holds. The conclusion is that the ``left"
off-critical algebra shares the same
form and same $g_r^{s_1,s_2}$ as the conformal $W_\infty (\widehat G)$-algebra
(note that $0\le m\le 2s-1$). The same is true for the ``right" algebra.

{}From the explicit form (4.6) of $ \oversome{(\!\!-)}\,L^{\,\,ij(2s)}_{-m},
 \oversome{(\!\!-)}\,Q^{\,\,ij(2s-1)}_{-m}$ one can easily see that ``left" and
``right" algebra do not commute. The problem of computing the structure
constants $\overline g_r^{s_1,s_2}$ of the mixed algebra ($m_1<m_2$):
$$
\eqalign{
\left[ L^{ij(2s_1)}_{-m_1},\overline L^{kl(2s_2)}_{m_2}\right]&=
(m^2)^{m_1}\sum ^{s_1+s_2-m_1-2}_{r=0}\left\{ \overline
g_{2r}^{2s_1,2s_2}\left(\delta\circ \overline L_{-m_2+m_1}^{2(s_1+s_2-m_1-1-r)}
\right)^{ijkl}\right.\cr
&+\left.\overline g_{2r-1}^{2s_1,2s_2}\left(\delta\circ \overline Q_{-m_2+m_1}
^{2(s_1+s_2-m_1-r)-1)}\right)^{ijkl}\right\}\cr}
$$

$$
\eqalign{
\left[ Q^{ij(2s_1-1)}_{-m_1},\overline Q^{kl(2s_2-1)}_{m_2}\right]&=
(m^2)^{m_1}\sum ^{s_1+s_2-m_1-2}_{r=0}\left\{ \overline
g_{2r}^{2s_1-1,2s_2-1}\left(\delta\star \overline L_{-m_2+m_1}^{2(s_1+s_2-m_1
-r)-1}\right)^{ijkl}\right.\cr
&+\left.\overline g_{2r-1}^{2s_1-1,2s_2-1}\left(\delta\star \overline Q_{-m_2+
m_1}^{2(s_1+s_2-m_1-r)-1)}\right)^{ijkl}\right\}\cr}
$$
%where
%$$\eqalign{\left[\delta\star {\cal Y}\right]^{ijkl}&=
%\delta^{ik}{\cal Y}^{jl}
% + \delta^{jl}{\cal Y}^{ik} - \delta^{il}{\cal Y}^{jk} -
%\delta^{jk}{\cal Y}^{il}\cr
%\left[\delta\circ {\cal Y}\right]^{ijkl}&=
%\delta^{ik}{\cal Y}^{jl}
% + \delta^{jl}{\cal Y}^{ik} + \delta^{il}{\cal Y}^{jk} +
%\delta^{jk}{\cal Y}^{il}\cr }$$
etc. is more complicated. One can apply in principle the method\ref{2} used for
$W_\infty(V)$-algebra in the case of $\widetilde W_\infty(\widehat G_n)$ as
well,
but up to now the problem of the computation of the mixed structure constants
$g_r^{s_1s_2}(n_1,n_2)$ is still open.
\vskip 2cm
\penalty-400

\centerline {{\bf 5.} Further generalizations}
\nobreak
\vskip .5cm
\nobreak

 The fact that the $O(n)$-massive fermionic action (1.1) has  ${\rm Vir}
\subset
\!\!\!\!\!\! \times \widehat{GL}(n,R)_{{\rm mod} 2}$ and $\widetilde W_\infty
(\widehat G_n)$ as algebras of symmetries leads to a natural question: {\it
whether
one can find similar infinite dimensional algebras studying more general
systems
of free massive fields}: $N$ fermions and $M$ bosons taken in appropriate
representations of the internal group $G_n$. The purely fermionic case is
straightforward generalization of the $n$-Majorana fermions in vector
representation of $O(n)$ we have  described above. For example all the
conserved charges of the massive fermions in the fundamental representation of
$A_N=SU(N+1)$ are generated by the conserved tensors (2.8) (written now for the
massive fermions $\psi^\alpha(z,\overline z)$). The case of massive bosons in
the vector representation of $O(n)$:
$$
S={1\over 2}\int d^2z \left( \partial \varphi^i\overline \partial \varphi^i +
m^2
\varphi^i\varphi^i\right) \eqno(5.1)
$$
requires certain modifications in the construction of the conserved tensors:
$$
\eqalign{
T^{ij}_{2s}&=\partial \varphi^i\partial ^{2s-1}\varphi^j + \partial
\varphi^j\partial
^{2s-1} \varphi^i\cr
J^{ij}_{2s-1}&= \varphi^i\partial ^{2s-1}\varphi^j -  \varphi^j\partial
^{2s-1} \varphi^i\quad .\cr}\eqno(5.2)
$$
Using the equation of motion one can easily obtain the corresponding
conservation laws:
$$
\eqalign{
&\overline \partial T^{ij}= \partial \theta ^{ij}\quad ,\quad \theta
^{ij}=m^2\varphi^i\varphi^j\cr
&\overline \partial J^{ij} + \partial \overline J^{ij} = 0 \quad ,\quad
\overline J^{ij} = \varphi^i\overline \partial \varphi^j - \varphi^j\overline
\partial
\varphi^i\cr
&\overline \partial T^{ij}_4 = \partial ^3\theta^{ij} + m^2 \partial T^{ij}\cr
&\overline \partial J^{ij}_3 = - \partial ^3\overline J^{ij} + m^2 \partial
J^{ij} \quad
\quad ,{\rm etc}\quad .\cr}\eqno(5.3)
$$

Due to the specific form of the r.h.s. of (5.3) $T^{ij}_{2s}$ and
$J^{ij}_{2s-1}$ do satisfy the criterion of ref.[2] of existence of new
noncommuting charges. The constructions of the conserved charges are similar to
the fermionic ones and as a consequence they span the same algebra $\widetilde
W_\infty (G_n)$.

Having a system of massive fermions and bosons one could expect larger
symmetries which mix the fermionic and bosonic degrees of freedom, i.e.
supersymmetric generalization $S\widetilde {W}_\infty(G_n)$ of the $\widetilde
W_\infty (G_n)$. The simplest case is of one Majorana fermion $\psi(z,\overline
z),\overline \psi(z,\overline z)$ and one boson $\varphi(z,\overline z)$. We
can
take the  conserved tensors in the form:
$$
\eqalign{
T_{2s}&=\psi\partial ^{2s-1}\psi + \partial \varphi\partial ^{2s-1}\varphi\cr
G_{2s-1/2}&=\psi\partial ^{2s-1}\varphi\quad ,\quad \overline G_{2s-1/2}=
\overline
\psi\, \overline \partial ^{2s-1}\varphi\quad .\cr}\eqno(5.4)
$$

The corresponding conservation laws have specific form with higher derivatives
in the r.h.s., which allows the construction of ``higher momenta" conserved
charges. For example:
$$
\eqalign{
\overline \partial G_{3/2}&=\partial \Theta\quad ,\quad \Theta=m\overline
\psi\varphi\, ,\, \overline \Theta =m \psi\varphi\cr
\partial \overline G_{3/2}&=-\overline \partial\,\, \overline \Theta\cr
\overline \partial G_{7/2} &= \partial ^3\Theta + m\partial ^2\overline \Theta
+ m^2 \partial G_{3/2}\quad ,\cr}
$$
etc. and therefore we can construct an infinite set of supersymmetric charges
$G^{(2s-1)}_{-k}$ (and $\overline G^{(2s-1)}_{-k}$):
$$
\eqalign{
{\cal G}_{-1/2}&=\int G_{3/2}dz - \int \Theta d\overline z\cr
{\cal G}_{-3/2}^{(3)}&=
\int G_{7/2}dz - \int \left( \partial ^2\Theta  + m\partial
\overline \Theta  + m^2G_{3/2}\right)d\overline z\cr
{\cal G}_{-1/2}^{(3)}&=
\int \left( z G_{7/2} + m^2\overline z G_{3/2}\right) dz -
\int \left( \partial ^2 z \Theta  - 3 \partial \Theta + m\partial
( z \overline \Theta ) - 2m \overline \Theta + m^2 z G_{3/2}\right)
d\overline z\cr}
$$
etc. Together with the conserved ``momenta" $L^{(2s)}_k$ of $T_{2s}$ they span
the $N=1$ supersymmetric analog of the $W_\infty(V)$ algebra [2]. By similar
constructions considering say three fermions and three bosons we can derive the
off-critical supersymmetric analog of $W_\infty(A_1)$, i.e. the current
superalgebra $\widetilde{SW}_\infty (A_1)$.

It becomes clear from this short discussion that having at hands free massive
fermions and bosons one can construct large class of $N$-supersymmetric and
supercurrent off-critical $W_\infty(G_n)$ algebras.

\vskip.5cm

\noindent {\bf Acknowledgement} This work of E.A. and M.C.B.A. has been
partially supported by CNPq, while that of G.S. and M.S. has been supported by
FAPESP.

\penalty-500
\vskip1cm
\penalty-500
\centerline{\bf References.}
\nobreak
\vskip.5cm
\nobreak
\refer[[1]/H. Itoyama, H.B. Thacker, {\it Nucl. Phys. } {\bf B320} (1989)541.]
\refer[[2]/G. Sotkov, M. Stanishkov IFT-preprint 001/93.]
\refer[[3]/L. Bonora, Y.Z. Zhang, M. Martellini, {\it Int. J. Mod. Phys.} {\bf
A6} (1991)1617.]
\refer[[4]/V. G. Knizhnik, A. B. Zamolodchikov, {\it Nucl. Phys. } {\bf B247}
(1984)83. ]
\refer[[5]/E. Witten {\it Commun. Math. Phys.} {\bf 92} (1984)455. ]
\refer[[6]/ C.N. Pope, X. Shen, L.J. Romans {\it Nucl. Phys. } {\bf B339}
(1990)191. ]
\refer[[7]/A. A. Belavin, A. M. Polyakov, A. B. Zamolodchikov {\it Nucl. Phys.
} {\bf B241} (1984)333.]
\refer[[8]/P. Furlan, G. Sotkov, I. T. Todorov, {\it Rivista del Nuovo Cimento}
{\bf 12} (1989)1.]
\refer[[9]/C.N. Pope, L.J. Romans, X. Shen, Phys. Lett. {\bf B}242 (1990)101.]
\refer[[10]/P. Gepner, E. Witten, Nucl. Phys. {\bf B}287(1987)493.]

\end